\documentclass[preprintnumbers,nofootinbib, amsmath,amssymb, prd,aps,superscriptaddress,twocolumn,floatfix]{revtex4-2}

\usepackage{graphicx}
\usepackage{amsmath}
\usepackage{natbib}
\usepackage[caption=false]{subfig}
\usepackage{siunitx}
\usepackage{placeins}
\usepackage{color}
\usepackage{standalone}
\usepackage{dcolumn}
\usepackage{tensor}
\usepackage{bm}
\usepackage{microtype}
\usepackage{etoolbox}
\usepackage{amssymb}
\usepackage{mathrsfs}
\usepackage{accents}
\usepackage[normalem]{ulem}
\usepackage{soul} 
\usepackage[dvipsnames]{xcolor}
\usepackage[colorlinks,urlcolor=NavyBlue,citecolor=NavyBlue,linkcolor=NavyBlue,pdfusetitle]{hyperref}
\usepackage[all]{hypcap}
\usepackage[inline]{enumitem}
\usepackage[utf8]{inputenc}
\usepackage{xspace}
\usepackage[printonlyused, nolist]{acronym}
\usepackage{lineno}

\newcommand{\irm}{{\rm i}}

\begin{document}

\title{The response of the Moon to gravitational waves}

\author{Xiaoming Bi}
\email{bixm20@lzu.edu.cn}
\affiliation{Lanzhou University (LZU), 730000 Lanzhou, Gansu, China}
\affiliation{Gran Sasso Science Institute (GSSI), I-67100 L'Aquila, Italy}

\author{Jan Harms}
\affiliation{Gran Sasso Science Institute (GSSI), I-67100 L'Aquila, Italy}
\affiliation{INFN, Laboratori Nazionali del Gran Sasso, I-67100 Assergi, Italy}

\begin{abstract}
The response of the Moon to gravitational waves (GWs) is used by some of the proposed lunar GW detectors like the Lunar Gravitational-wave Antenna (LGWA) to turn the Moon into an antenna for GWs. The deep connection between the lunar internal structure, its geophysical environment and the study of the Universe is intriguing, but given our limited understanding of the Moon today, it also makes it very difficult to predict the science potential of lunar GW detectors accurately. Lunar response models have been developed since the Apollo program, but there is evidence coming from seismic measurements during the Apollo missions that the models are not good enough and possibly underestimating the lunar GW response especially in the decihertz frequency band. In this paper, we will provide an extension of Freeman Dyson's half-space model to include horizontally layered geologies, which allows us to carry out computationally efficient calculations of the lunar GW response above 0.1\,Hz compared to the normal-mode simulations used in the past. We analyze how the results depend on the values of geometric and elastic parameters of the layered geological model, and we find that modifications of the geological model as required to explain Apollo seismic observations can boost the lunar GW response. 
\end{abstract}

\maketitle

\section{Introduction}
The Moon is an ideal platform for gravitational-wave (GW) detectors to observe signals below the frequency band accessible to current and future terrestrial detectors \cite{Branchesi2023}. The Moon is seismically exceptionally quiet and at its permanently shadowed regions, it is exceptionally cold and thermally stable. Upper limits set to the spectrum of the lunar seismic background with Apollo seismometers are up to 4 orders of magnitude weaker in amplitude than seismic spectra on Earth between 0.1\,Hz and 1\,Hz \cite{CoHa2014c}, and models predict that the background might be more than 7 orders of magnitude weaker than on Earth \cite{LoEA2009,HaEA2021a}. These close to ideal characteristics of the Moon and today's revived interest in lunar science and exploration have led to new proposals for lunar GW detectors originally presented to space agencies in 2020 \cite{Branchesi2023,HaEA2021a,JaLo2020}, with additional studies in subsequent years \cite{ASEA2021,Li2023lunar}. The aim is to realize GW detections in the band 1\,mHz to a few Hz with special focus on the decihertz band, which will remain unobserved by the LISA mission \cite{colpi2024lisa} as well as current and future terrestrial GW detectors \cite{ET2020,EvEA2021}.

Among the proposed concepts is the Lunar Gravitational-wave Antenna (LGWA), which requires high-performance accelerometers to observe the Moon's surface vibrations caused by GWs \cite{HaEA2021a}. The lunar response to GWs is a crucial quantity to estimate the sensitivity of LGWA to GWs, which will be limited by instrument-noise of the accelerometer \cite{vHeEA2023} and by the lunar seismic background \cite{Har2022a}. So far, two approaches were used to model the lunar GW response. The first is based on a normal-mode analysis, which is most effective for response calculations from 1\,mHz to a few 10\,mHz \cite{Ben1983,CoHa2014b,HaEA2021a,Branchesi2023}. The normal-mode analysis is based on a model of the internal structure of the Moon as given for example in Garcia et al \cite{GaEA2019}. It is computationally very challenging to extend such models up to a few Hz, and the normal-mode analysis would become impracticable if deviations from radial symmetry had to be accounted for such as topography and heterogeneous geology. These features are expected to be important for the lunar response modeling in the decihertz band.

The second approach is a phenomenological model based on Dyson's homogeneous half-space response calculations \cite{Dys1969}. The current recipe adopted for the decihertz lunar GW response and decihertz LGWA sensitivity predictions is to calculate resonant corrections to the simple homogeneous half-space model \cite{cozzumbo2023opportunities}. These resonances are assumed to be produced by some geological features with an effective Q-factor of a few 100, but without underlying physical model. This approach is unsatisfying, but it is justified by how site effects are generally modeled and taking into consideration that effective Q-factors of a few 1000 were inferred from active studies and moonquake observations during the Apollo missions \cite{LaEA1970,GaEA2019}. Nevertheless, it is crucial to provide a proper decihertz response model, which is the aim of this study.

In section \ref{sec:normalmode}, we revisit the normal-mode formalism to facilitate analytical comparisons with the new horizontally layered model presented in section \ref{sec:horizontally}. The new model is analyzed in section \ref{sec:variations} in terms of parametric variations, and we briefly discuss the options that would explain the high Q-values inferred from moonquake observations and provide a simple representation of the involved physics to be able to calculate the lunar response consistent with high Q-values. We then use the new response model to calculate the corresponding LGWA sensitivity curve in section \ref{sec:sensitivity}.

\section{Normal-mode method to calculate the lunar GW response}
\label{sec:normalmode}
The preferred method so far to calculate the lunar response to GWs was to use the normal-mode formalism initially adapted for this purpose by Ben-Menahem \cite{Ben1983}. In this section, we briefly review the normal-mode method so that it can later be compared with the equations of the Dyson model.

Any displacement field $\vec\xi(\vec r,t)$ of the Moon can be written as a sum over normal modes $\vec\chi_N$ with amplitude $A_N$:
\begin{equation}
    \vec\xi(\vec r,t)=\sum\limits_N A_N(t)\vec\chi_N(\vec r\,).
    \label{eq:displ}
\end{equation}
The index $N$ can generally stand for several indices such as $n$, $l$, $m$ in the formalism of spherical harmonics. In this paper, we choose the following normalization of the normal modes
\begin{equation}
    \int\limits_{V}{\rm d}^3r\,\rho(\vec r\,)\vec\chi_{N'}(\vec r\,)^*\cdot\vec\chi_N(\vec r\,)=M\delta_{NN'},
\label{eq:normalization}
\end{equation}
where $\rho(\vec r\,)$ is the mass density of the Moon at position $\vec r$, and $M$ is the mass of the Moon. Note that the normal modes are without unit in this normalization, while the amplitude $A_N$ has units of displacement. 

There are two ways to describe the coupling of a GW with the Moon. The first is to consider the Moon as a collection of infinitesimal mass elements, and to describe the effect of the GW as a tidal force field acting on these masses. This force field drives the normal-mode amplitude according to
\begin{equation}
\begin{split}
    &\ddot A_N(t)+\frac{\omega_N}{Q_N}\dot A_N(t)+\omega_N^2A_N(t)=\\
    &\qquad\frac{1}{2M}\int{\rm d}^3r\,\rho(\vec r\,)\vec\chi_N(\vec r\,)^*\cdot\ddot{\mathbf h}(\vec r,t)\cdot\vec r. 
\end{split}
\end{equation}
Each normal mode represents a harmonic oscillator at frequency $\omega_N$ subject to viscous damping described by the quality factor $Q_N$. The GW is represented by the spatial ($3\times 3$) components $\mathbf h(\vec r,t)$ of the metric perturbation. We can typically neglect the variation of the metric perturbation over the extent of the Moon since the lengths of the GWs is much longer than the diameter of the Moon at the frequencies of interest.

When considering the measurement of the surface displacement with a seismometer, one needs to keep in mind that the force field also acts on the suspended proof mass of the seismometer. This means that once the displacement $\vec\xi(\vec r_0,t)$ of the Moon's surface is calculated at the position $\vec r_0$ of the seismometer, the term $\mathbf h(\vec r_0,t)\cdot\vec r_0/2$ needs to be subtracted. 

The second, in many ways more convenient approach is to describe the action of a GW on the Moon in transverse-traceless gauge, where the coupling happens with the gradient of the shear modulus, $\nabla\mu(\vec r\,)$:
\begin{equation}
\begin{split}
    &\ddot A_N(t)+\frac{\omega_N}{Q_N}\dot A_N(t)+\omega_N^2A_N(t)=\\
    &\qquad-\frac{1}{M}\int{\rm d}^3 r\,\vec\chi_N(\vec r\,)^*\cdot{\mathbf h}(t)\cdot\nabla\mu(\vec r\,).
\end{split}
\label{eq:respTT}
\end{equation}
This approach was also taken by Dyson and Ben-Menahem \cite{Dys1969,Ben1983}, and its equivalence with the first approach was shown in \cite{Har2019}. In this case, there is no acceleration of the seismometer proof mass that would have to be accounted for: the displacement $\vec\xi(\vec r,t)$ calculated in this way directly describes the displacement signal of the seismometer! 

In order to make the connection with the seminal work of Ben-Menahmen more explicit, we now consider the simple case of a Moon with radially symmetric internal structure, i.e., $\rho(\vec r\,)=\rho(r)$ and $\mu(\vec r\,)=\mu(r)$. As was shown in \cite{Ben1983}, the response of toroidal modes to GWs is strongly suppressed, and we can focus on the spheroidal normal modes,
\begin{equation}
    \vec\chi_{nl}^{\,\rm S}(\vec r\,)=a_{nl}(r)\vec Y_2^m(\theta,\phi)+\sqrt{l(l+1)}b_{nl}(r)\vec\Psi_l^m(\theta,\phi),
    \label{eq:spheroidal}
\end{equation}
expressed in terms of the vector spherical harmonics
\begin{equation}
\begin{split}
    \vec Y_l^m(\theta,\phi) &= Y_l^m(\theta,\phi)\vec e_r,\\
    \vec\Psi_l^m(\theta,\phi) &= \frac{1}{\sqrt{l(l+1)}}r\nabla Y_l^m(\theta,\phi).
\end{split}
\end{equation}
The vector harmonics are normalized such that they form an orthonormal basis
\begin{equation}
\begin{split}
    \int{\rm d}\Omega\, \vec Y_{l'}^{m'}(\theta,\phi)\cdot\vec Y_l^m(\theta,\phi)^*=\delta_{ll'}\delta_{mm'},\\
    \int{\rm d}\Omega\, \vec\Psi_{l'}^{m'}(\theta,\phi)\cdot\vec \Psi_l^m(\theta,\phi)^*=\delta_{ll'}\delta_{mm'},
\end{split}
\label{eq:orthon}
\end{equation}
with products between different vector spherical harmonics vanishing. The normalization condition of the normal modes in equation (\ref{eq:normalization}) now reads
\begin{equation}
     \int\limits_0^{R}{\rm d}r\,r^2\rho(r)\left(a_{nl}^2+l(l+1)b_{nl}^2\right)=M,
\end{equation}

As a simple demonstration, we now use equation (\ref{eq:respTT}) to calculate the lunar response explicitly for a homogeneous Moon. We choose the GW tensor to take the form $\mathbf h(t)=h_0(t)(\vec e_x\otimes\vec e_x-\vec e_y\otimes\vec e_y)$, i.e., the GW is propagating along the $z$-axis, which aligns with the polar axis of the spherical coordinate system used in the previous equations, and we neglect the position dependence of the GW amplitude assuming that the Moon is much smaller than the length of the GW. Inserting this expression into the right-hand-side of equation (\ref{eq:respTT}), and substituting the normal mode by the expression in equation (\ref{eq:spheroidal}), we obtain for quadrupole modes $l=2$:
\begin{widetext}
\begin{equation}
\begin{split}
    &-\frac{1}{M}\int{\rm d}^3 r\,\vec\chi_N(\vec r\,)^*\cdot{\mathbf h}(t)\cdot\nabla\mu(r) = \frac{R^2\mu}{M}\int{\rm d}\Omega\,\vec\chi_N(R,\Omega)^*\cdot{\mathbf h}(t)\cdot\vec e_r \\
    &\quad=\frac{R^2\mu}{M}h_0(t)a_{n2}(R)\int{\rm d}\Omega\,Y_2^m(\theta,\phi)^*((\vec e_r\cdot\vec e_x)^2-(\vec e_r\cdot\vec e_y)^2)\\
    &\qquad +\frac{R^2\mu}{M}h_0(t)b_{n2}(R)\sqrt{6}\int{\rm d}\Omega\,\vec\Psi_2^m(\theta,\phi)^*\cdot(\vec e_x(\vec e_r\cdot\vec e_x)-\vec e_y (\vec e_r\cdot\vec e_y))\\
    &\quad=\frac{R^2\mu}{M}h_0(t)2\sqrt{\frac{2\pi}{15}}\left(a_{n2}(R)+3b_{n2}(R)\right),
\end{split}    
\end{equation}
\end{widetext}
where $R$ is the radius of the Moon, ${\rm d}\Omega={\rm d}\phi{\rm d}\theta\sin(\theta)$, and we made use of $\nabla\mu(r)=-\mu\,\delta(r-R)\vec e_r$ ($\vec e_r$ is the unit radial vector). The frequency-domain response can then be written
\begin{equation}
    \tilde A_n(\omega)=\frac{1}{2}\tilde h_0(\omega)L_n\frac{\omega_n^2}{\omega_n^2-\omega^2-{\rm i}\omega_n\omega/Q_n},
\label{eq:amphomo}
\end{equation}
where the effective baseline $L_n$ is given by
\begin{equation}
    L_n=\frac{3\beta^2}{\pi R\omega_n^2}2\sqrt{\frac{2\pi}{15}} \left(a_{n2}(R)+3b_{n2}(R)\right).
\end{equation}
We introduced $\beta^2=\mu/\rho$, where $\beta$ is the speed of seismic shear waves. A factor 2 was multiplied to the baseline since the two modes with $m=-2,2$ out of the 5 quadrupole modes respond equally to the GW, the other three modes do not respond. Instead, in a rotating or asymmetric lunar model, the modes can be split and the baseline must adopt the index $m$ as well. The expressions for the functions $a_{n2},\,b_{n2}$ can for example be found in equation (B14) in \cite{Lobo1995}, which also contains equations how to calculate the normal-mode frequencies $\omega_n$. Equation (\ref{eq:amphomo}) together with equation (\ref{eq:spheroidal}) can be inserted into equation (\ref{eq:displ}) to obtain the full GW response of the homogeneous Moon. 

Technically, the difficult step of the normal-mode model is to calculate the functions $a_{n2}(r)$, $b_{n2}(r)$ for arbitrary radially symmetric internal structure models together with the respective normal-mode frequencies and Q-values \cite{Woodhouse1988}. The numerical tool MINOS\footnote{https://igppweb.ucsd.edu/~gabi/rem.dir/surface/minos.html} was used in a past study  \cite{CoHa2014b}. The sum in equation (\ref{eq:displ}) is then carried out up to some order $n$ typically set by the frequency band of interest. In any case, it is computationally challenging to calculate the normal modes with high order $n$, and it is also not necessarily reasonable to push to higher frequencies with normal-mode simulations where heterogeneities of the Moon play a larger role. The results presented in \cite{Branchesi2023} include normal modes up to order $n=229$ with $f_{229}=\omega_{229}/(2\pi)$ between 0.1\,Hz and 0.2\,Hz.

\section{Dyson model of the lunar GW response for a horizontally layered geology}
\label{sec:horizontally}
The first model of the response of an elastic body to GWs was calculated by Dyson for a homogeneous half space \cite{Dys1969}. In this section, we will apply the same approach to calculate the GW response of a horizontally layered half space. 

The layered geology means that the elastic parameters only change at a discrete set of depths, i.e., the layer interfaces. Inside each layer, the medium is homogeneous (and isotropic) and the equation of motion inside each layer assume the simple form without GW term:
\begin{equation} \rho(\vec r\,) \ddot{\vec\xi}(\vec r,t)=(\lambda+\mu)\nabla\left(\nabla\cdot\vec\xi(\vec r,t)\right)+\mu\nabla^2\vec\xi(\vec r,t).
\end{equation}
However, the GW influences the boundary condition between layers and at the surface:
\begin{widetext}
\begin{eqnarray}
\vec\tau_{k+}-\vec\tau_{k-} + (\mu_k-\mu_{k+1})\vec n\cdot\mathbf h(\vec r_k,t) = 0,\label{eq:boundary1}\\
\vec\tau_{k+}=\lambda_k\vec n(\nabla\cdot\vec\xi(\vec r,t))\big|_{k+}+\mu_k\left(\nabla(\vec n\cdot\vec\xi(\vec r,t))\big|_{k+}+(\vec n\cdot\nabla)\vec\xi(\vec r,t)\big|_{k+}\right)\label{eq:boundary2}\\
\vec\tau_{k-}=\lambda_{k+1}\vec n(\nabla\cdot\vec\xi(\vec r,t))\big|_{k-}+\mu_{k+1}\left(\nabla(\vec n\cdot\vec\xi(\vec r,t))\big|_{k-}+(\vec n\cdot\nabla)\vec\xi(\vec r,t)\big|_{k-}\right)\label{eq:boundary3}\\
\vec\xi(\vec r,t)\big|_{k+}=\vec\xi(\vec r,t)\big|_{k-} \label{eq:boundary4}
\end{eqnarray}
\end{widetext}
where the index $k$ counts the layers, $\big|_{k\pm}$ means to evaluate a function at the interface either on its upper ($+$) or lower ($-$) side, and $\vec n$ is the unit vector normal to the horizontal interfaces. The vacuum above surface counts as additional layer $k=0$ with $\mu_0=0$. The continuity of displacement --- last row in equation (\ref{eq:boundary4}) --- only holds for solid-solid interfaces. At fluid-solid interfaces, only the displacement normal to the interface is continuous, while at the free surface, the continuity condition is not applied.

Solving the equations for the displacement field can be simplified by realizing that equation (\ref{eq:boundary1}) can only be fulfilled if the seismic waves produced by the interaction of the ground with GWs propagate vertically across the layered structure, i.e., in the direction of $\vec n$. This is because the GW amplitude is to a very good approximation the same across the interface (across the entire Moon, since we are interested in frequencies where the GW length is at least on the order of hundreds of thousands of kilometers). As a consequence, the seismic waves produced at the interfaces also need to have nearly the same amplitude across the interface, and this can only be achieved if they propagate (nearly) vertically.

We can therefore express the displacement field in each layer as a sum of an upwards and downwards traveling seismic wave
\begin{equation}
\vec\xi(\vec r,t)= \sum\limits_{p=\alpha,\beta} \vec\xi^{\rm d}_p(t)\mathrm{e}^{\irm k_pz} +\vec\xi^{\rm u}_p(t)\mathrm{e}^{-\irm k_pz},
\end{equation}
where the sum is over the two polarizations with $k_\alpha=\omega/\alpha$, $k_\beta=\omega/\beta$ being the wavenumbers of compressional and shear  waves, $\alpha,\,\beta$ being the speed of compressional and shear waves, $\vec\xi_\alpha(t)=(0,0,\xi_\alpha(t))$ and $\vec\xi_\beta(t)=(\xi_{x,\beta}(t),\xi_{y,\beta}(t),0)$, and $z$ is the coordinate along the vertical direction. We assume that there is no seismic wave incoming from below the lowest layer interface. It is also possible to simulate damping of the layer material by adding a small imaginary part to the wavenumbers, e.g., $k\rightarrow k(1+\irm/(2Q))$, where $Q$ is the medium's quality factor and generally depends on the wave polarization.

The solution for a homogeneous half space was first calculated by Dyson and takes the form \cite{Dys1969}
\begin{equation}
\begin{split}
&\xi_x=-\irm \dfrac{1}{k_\beta}h_{xz},\, \xi_y=-\irm \dfrac{1}{k_\beta}h_{yz},\\
&\xi_z=\irm\dfrac{\mu}{\lambda+2\mu}\dfrac{1}{k_\alpha}h_{zz}=\irm\dfrac{\beta}{\alpha}\dfrac{1}{k_\beta}h_{zz}.
\end{split}
\end{equation}
This model predicts a simple $1/\omega$ dependence of the lunar GW response. According to the last equation, we can formally define the baseline $L_{x,y}=\lambda_\beta/\pi$ for measurements of the horizontal displacement and $L_z=(\beta/\alpha)\lambda_\beta/\pi$ for measurements of the vertical displacement, where $\lambda=2\pi/k$ is the seismic wavelength. Since the speed $\alpha$ of compressional waves is always larger than the speed $\beta$ of shear waves, the GW response is stronger in the horizontal than in the vertical. 

\begin{table}[ht!]
\centering
\begin{tabular}{|l|r|r|r|r|r|r|}
\hline
Layer & 0 & 1 & 2 & 3 & 4 & {\color{magenta}5} \\\hline
Depth [km] & above surface & 0.00  & 1.00  &12.00 &28.00 & {\color{magenta}1317.1} \\\hline
$\alpha$ [km/s] & 0.00   & 1.00  &3.20  &5.50   &7.68  & {\color{magenta}4.00}    \\\hline
$\beta$ [km/s] & 0.00   & 0.50  &1.80  &3.30   &4.41  & {\color{magenta}0.00}  \\\hline
$\rho$ [g/cm$^3$] & 0.00  & 2.60  &2.60  &2.60   &3.34  & {\color{magenta}4.16}    \\\hline     
\end{tabular}
\caption{Reference geological model used in this paper, which is a simplification of model M2 in \cite{GaEA2019}. The depth value refers to the depth where the layer starts. The last layer formally represents a fluid core of the Moon. Unless explicitly mentioned in a figure caption, it is not included in our simulations.}
\label{tab:refmodel}
\end{table}
It is interesting to compare the result of the horizontally layered model with the normal-mode model. We do not expect a good match at frequencies where the finite size and spherical shape of the Moon are important. However, at frequencies above 0.1\,Hz, where the seismic wavelength is more than an order of magnitude smaller than the radius of the Moon, the two approaches should give similar results. In order to compare with the normal-mode result in \cite{Branchesi2023}, we will define a layered geology based on one of the layered models in Garcia et al \cite{GaEA2019}. The models in \cite{GaEA2019} are not in perfect agreement, but for GW response calculations, the relatively small discrepancies do not have an important impact. In table \ref{tab:refmodel}, we summarize the relevant parameters, which we will use as our reference model throughout this paper. Layer $k=5$ formally represents a fluid core of the Moon, which is not considered part of our reference model since the deep structure, albeit important, requires normal-mode simulations for accurate modeling.

\begin{figure}[ht!] 
  \centering 
  \includegraphics[width=0.99\columnwidth]{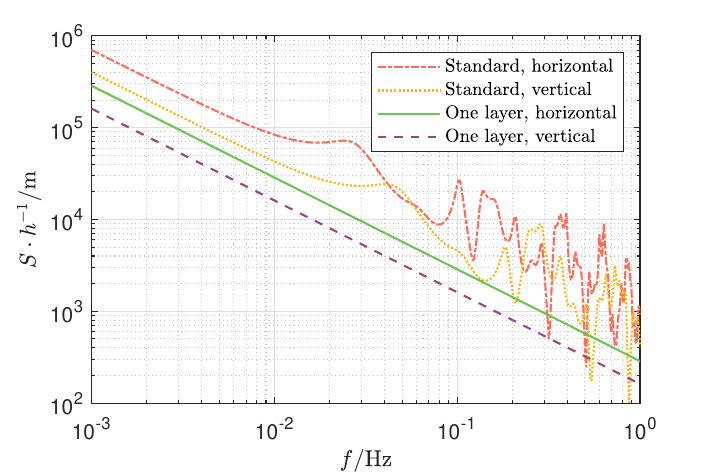} 
  \caption{Lunar GW response model considering only a shallow horizontally layered geology based on the parameter values in table \ref{tab:refmodel} together with the prediction for a homogeneous half-space model ("one layer").}
  \label{fig:refresponse}
\end{figure}
The result is shown in figure \ref{fig:refresponse}. In addition to the reference model based on the values in table \ref{tab:refmodel}, we show the prediction of the homogeneous half-space model using the values of layer 1 for the entire half space. Including only the shallow layers of the reference mode, the simulation predicts a lower (by about a factor 3) GW response around 0.1\,Hz compared to the normal-mode model in \cite{Branchesi2023}. This can be attributed to the presence of additional and sharper peaks in the normal-mode model in this part, which are produced by the finite size of the Moon and the liquid core, i.e., seismic waves propagating towards the center of the Moon and contributing to an energy loss in the shallow layered geology, can in reality come back and enhance the lunar GW response. We therefore conclude that the two models produce consistent predictions of the lunar GW response.

\section{Variations of the reference response model to produce the high-Q features in lunar seismic observations}
\label{sec:variations}
A gravitational wave produces seismic waves in a very peculiar way and distinct from any other known seismic source. The dominant coupling happens at all of the locations where strong changes of the shear modulus occur. At these locations, GWs drive emission of seismic waves along both directions parallel to the shear-modulus gradient, and this emission is coherent over the full extent of the Moon. Among the strongest emitters are the surface and solid-fluid interfaces characterized by a maximal change of the shear modulus (one side having vanishing shear modulus). However, far more important than the emission mechanism is what happens to seismic waves in a layered geology. In this paper, we adopt the half-space model, which means that a seismic wave transmitting into the deepest layer of the model is lost. Even for a more realistic spherical model of the Moon, it is reasonable to assume that at high enough frequencies, scattering and dissipation make sure that above some frequency, seismic energy emitted towards deep layers will not return in significant amounts. However, what if there is something preventing seismic energy to be transmitted into deep layers? In the following, we will show results from parametric variations of the shallow model and in the end arrive at a scenario with greatly enhanced GW response and Q-values consistent with the Apollo seismic observations.

\begin{figure}[htb] 
  \centering 
  \includegraphics[width=0.99\columnwidth]{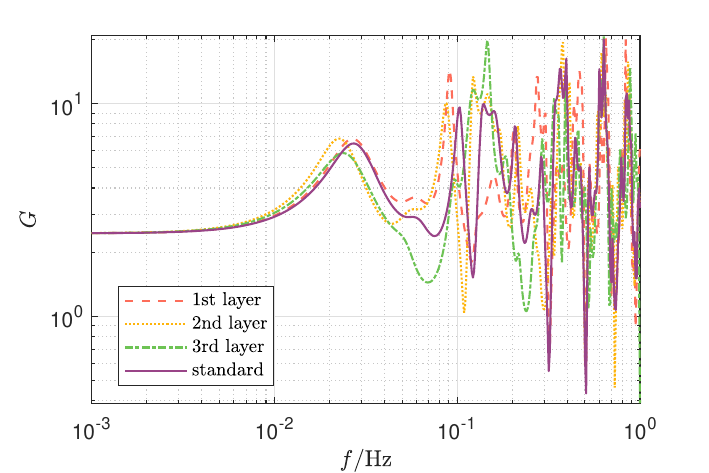} 
  \caption{Response changes when the thickness of one of the first three layers is increased by 33\%. Responses are normalized with respect to the homogeneous half-space model.}
  \label{fig:varythickness}
\end{figure}
The first study is the variation of the thickness of the first three layers ($k=1,2,3$ in table \ref{tab:refmodel}). In figure \ref{fig:varythickness}, we show the shallow-layered response model for the horizontal displacement normalized by the homogeneous half-space model (with parameter values of layer $k=1$ in table \ref{tab:refmodel}); both displayed in figure \ref{fig:refresponse}. The other curves show how the response changes if the thickness of the respective layer is increased by a factor 1.33. As one would expect, the frequencies of some of the peaks shift slightly under thickness variations. In addition, peak response changes by more than a factor 2. Changes in peak response through thickness variations would not be possible in a two-layer structure. It is a consequence of changing interference conditions between seismic waves resonating in different layers and then propagating to the surface. In this simulation, the ground medium is assumed to be dissipation free, which means that energy loss occurs exclusively by transmission into the lowest layer. Large reduction of energy loss due to layer thickness variations is not possible.

\begin{figure}[ht!] 
  \centering 
  \includegraphics[width=0.99\columnwidth]{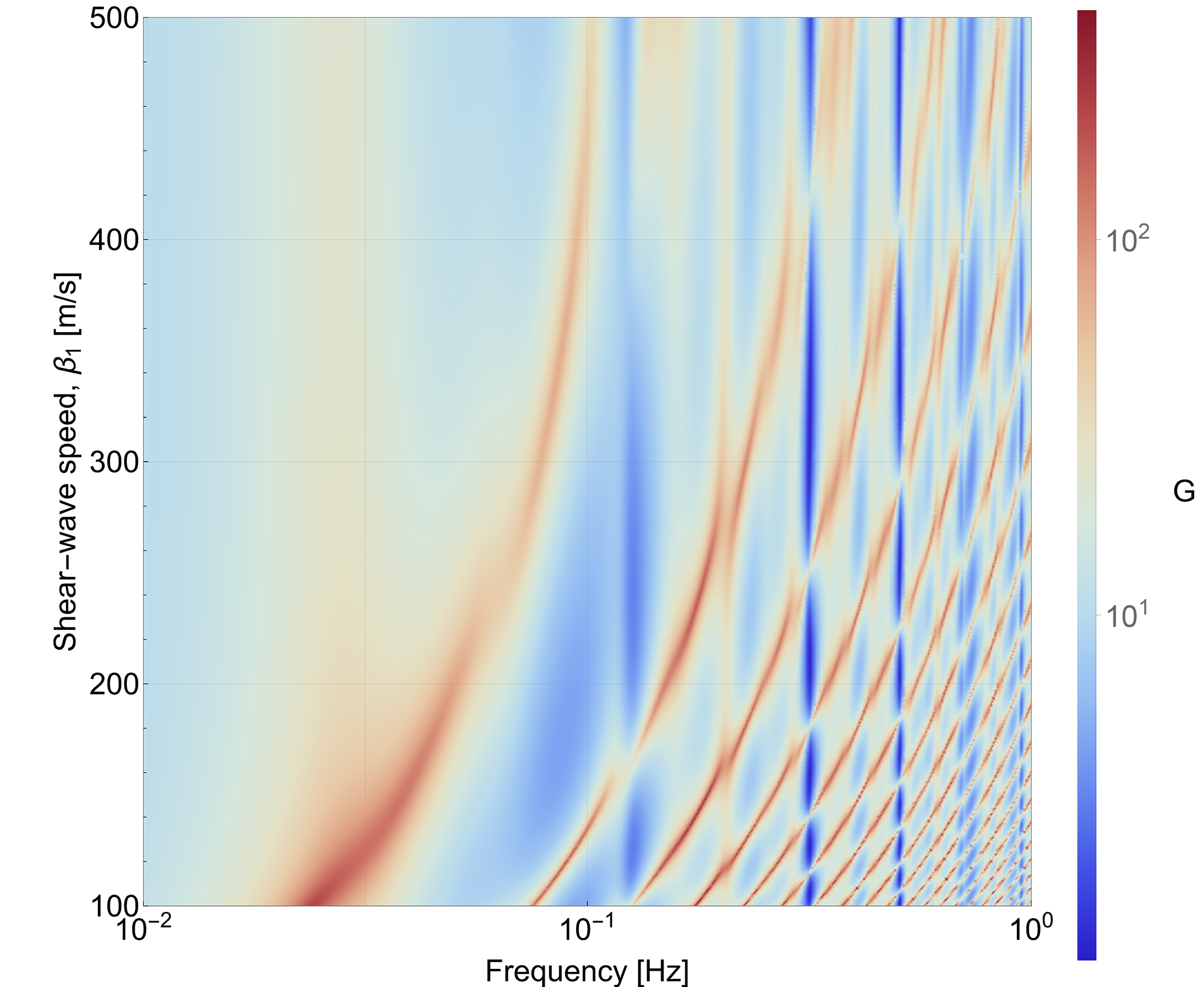} 
  \caption{Dependence of lunar GW response on the shear-wave speed of the first layer ($k=1$). The seismic speed of the reference model is $\beta_1=500$\,m/s.}
  \label{fig:varyspeed}
\end{figure}
A more important parameter is the shear-wave velocity $\beta$. In figure \ref{fig:varyspeed}, we show the change of response under variations of $\beta_1$ from 100\,m/s to the nominal shear-wave speed of 500\,m/s in layer $k=1$. It can be seen that the slower the speed in the surface layer, the stronger the GW response. There are three different contributions changing with $\beta_1$. The first is the emission of seismic waves from the surface, which becomes weaker with decreasing $\beta_1$ (since the shear modulus step at the surface becomes smaller). Second, the seismic waves emitted from the first interface become stronger with decreasing $\beta_1$, because the shear-modulus step becomes larger towards the layer $k=2$. These two effects would not necessarily point towards an overall enhancement of the response with decreasing speed. The third effect is due to the increase of the contrast across the first interface (between layers $k=1$ and $k=2$), which reduces the transmission into the lower layers and increases the resonant amplification within the surface layer $k=1$. This is in fact the explanation of why the GW response is larger when the shear-wave speed in the surface layer becomes slower. We now see that the ground response is resonantly amplified by up to a factor $100$. In conclusion, changes of the shear-wave speed of the surface layer can have an important impact on the lunar GW response and the slower the speed the stronger is the response. 

\begin{figure}[ht!]
\includegraphics[width=0.99\columnwidth]{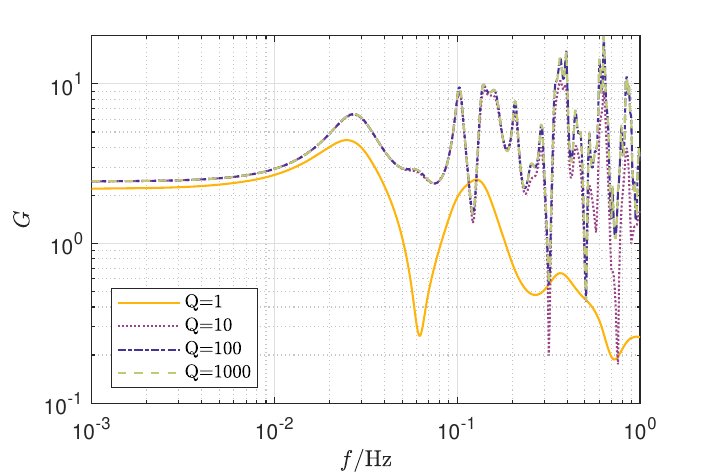}
\includegraphics[width=0.99\columnwidth]{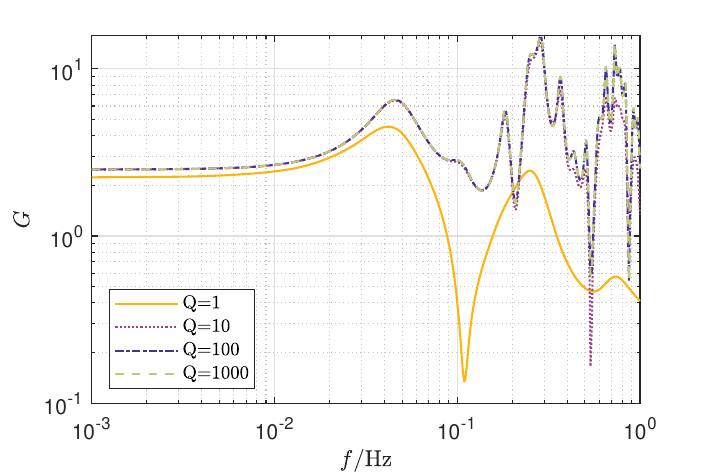}
\caption{Changes in lunar GW response under variations of the Q-value of the layer material (same Q-value for all layers). Upper plot: horizontal response, lower plot: vertical response.} 
\label{fig:qvalue}
\end{figure}
The next study is of the effect of the Q-value of the layer material on the lunar GW response. The results are shown in figure \ref{fig:qvalue}. Accordingly, the Q-value must be very low to have a significant effect on the response below 1\,Hz. Only for $Q=1$, the response changes dramatically and only above 10\,mHz. One can see that the height of peaks close to 1\,Hz is decreased significantly for $Q=10$, but not for $Q=100$. This indicates that the effective Q-value of the layered structure limited by energy loss into the lowest layer is well below $Q=100$. This is also clear from looking at the amplification factor, i.e., the peak height above the off-resonance response. If the resonances due to the layered structure were able to sustain Q-values in the thousands, then the material quality might set important limits to the GW-response amplifications. Note that the Q-value would produce stronger limitations if deeper layers played a role so that the propagation distance became longer.

\begin{figure}[ht!]
\includegraphics[width=0.99\columnwidth]{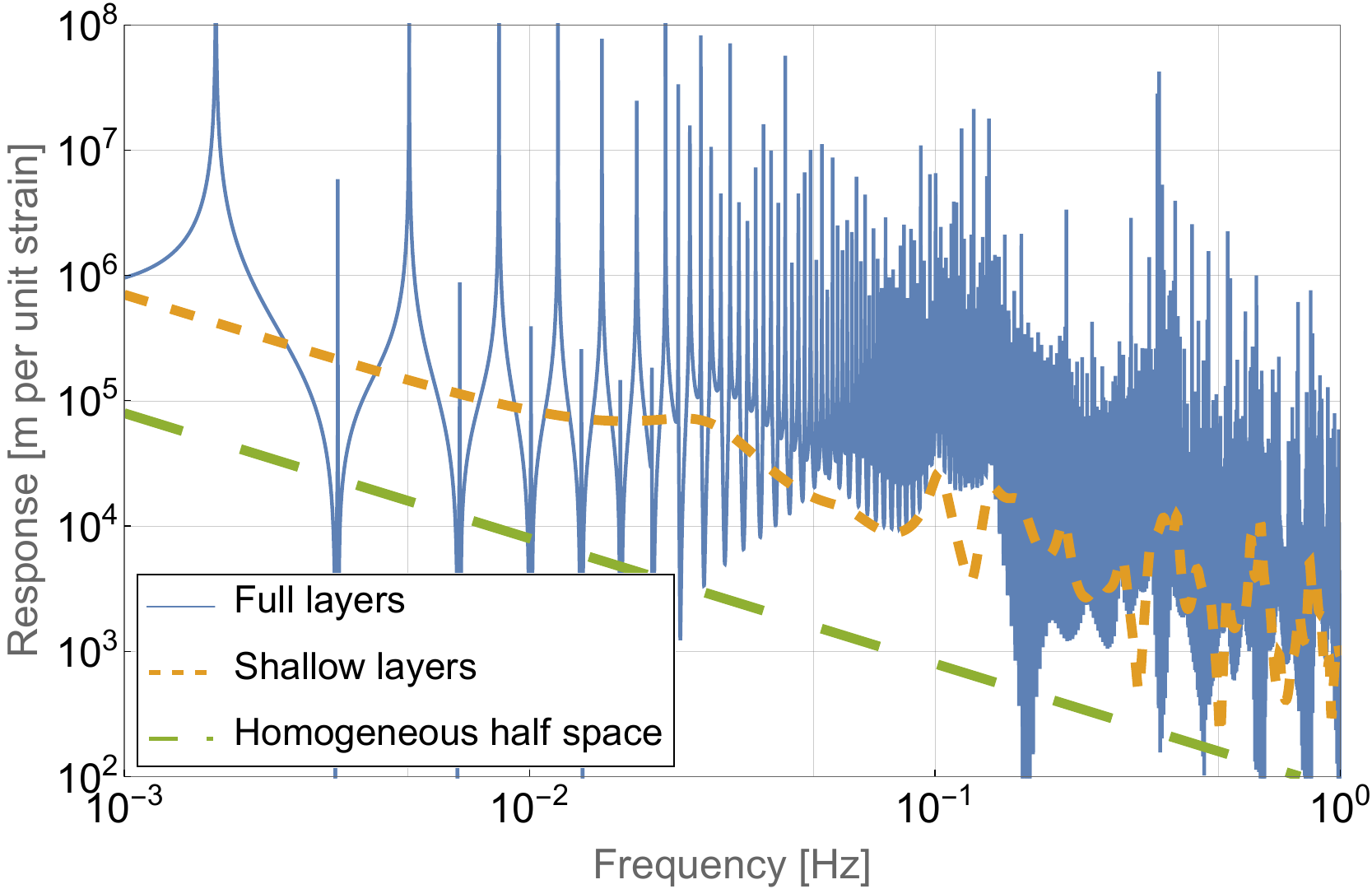}
\includegraphics[width=0.99\columnwidth]{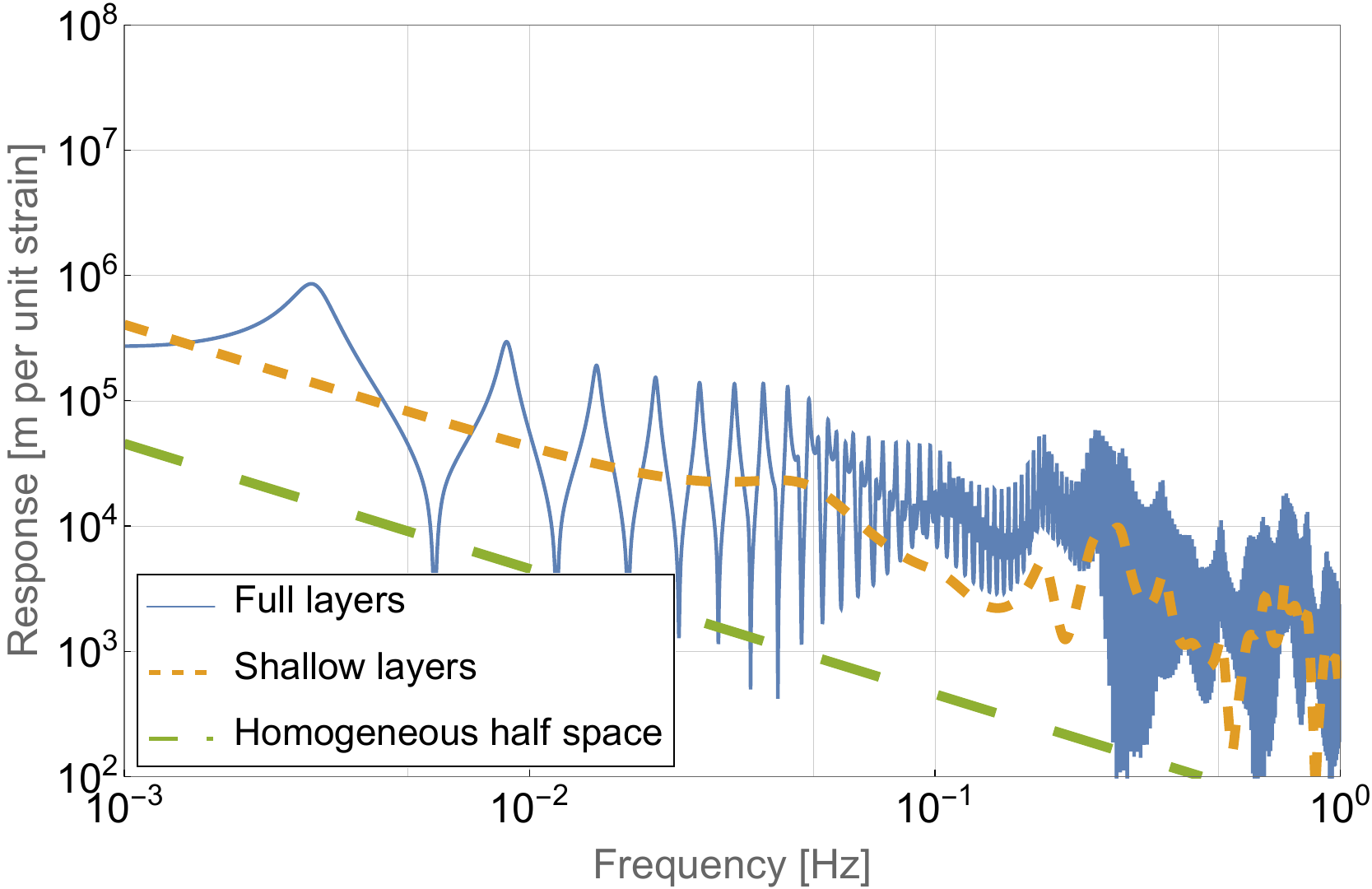}
\caption{In this simulation, the deep fluid layer $k=5$ is included and plotted together with the GW response of the shallow-layered model and of the homogeneous half-space model. Upper plot: horizontal response, lower plot: vertical response.}.
\label{fig:fluidresp}
\end{figure}
The next simulation includes the fluid layer $k=5$ from table \ref{tab:refmodel}. We do this exercise to illustrate important aspects of the lunar GW response, but we cannot expect that a fluid layer below a depth of 1370\,km corresponding to the depth of the suspected fluid core, leads to accurate results in a horizontally layered geological model. With such a fluid layer, downwards propagating shear waves are fully reflected from the solid-to-fluid interface (at normal incidence). This means that a fluid layer in a horizontally layered structure at any depth eliminates transmission loss into deeper layers, which leads to the sharp peaks in plot (a) in figure \ref{fig:fluidresp}. The height of these peaks is limited by the Q-value of the layer materials. Instead, compressional waves can propagate into a fluid layer, which means that energy loss due to transmission into the lowest layer is still possible, and the GW response in vertical displacement does not show the sharp peaks of the horizontal response. In fact, the deep fluid layer has a rather mild impact on the GW response in vertical direction; at least above a few 10\,mHz. 

Responses in horizontal and vertical direction for the model with deep fluid layer both show suppression below 1\,mHz with respect to the shallow-layer model. This is because at these low frequencies the long seismic waves emitted from the solid-to-fluid interface interfere destructively with the seismic waves propagating downwards from the upper layers. The horizontal response in the limit $\omega\rightarrow 0$ with $M$ layers and layer $k=M$ being a fluid ($\mu_M=0$) is governed by
\begin{equation}
    \xi_x=-\dfrac{1}{2}\dfrac{\sum_{k=1}^{M-1}\rho_k (h_{k+1}^2-h_k^2)}{\sum_{k=1}^{M-1}\rho_k (h_{k+1}-h_k)}h_{xz}=-\dfrac{1}{2}L_0h_{xz},
\end{equation}
which yields a baseline $L_0\approx 1323$\,km with the parameter values of table \ref{tab:refmodel}. Also the normal-mode response model has a low-frequency limit converging to a baseline that is equal to the lunar radius of 1737\,km.

In any case, an important conclusion is that a fluid layer can strongly enhance the lunar GW response. Any deeper layer with low shear-wave speed would have such an effect, and one might even speculate if dense fracture networks could lead to a significant energy return and enhance the GW response. These results provide another motivation to monitor horizontal surface displacement as proposed for LGWA instead of vertical displacement.

\begin{figure}
   \centering \includegraphics[width=0.99\columnwidth]{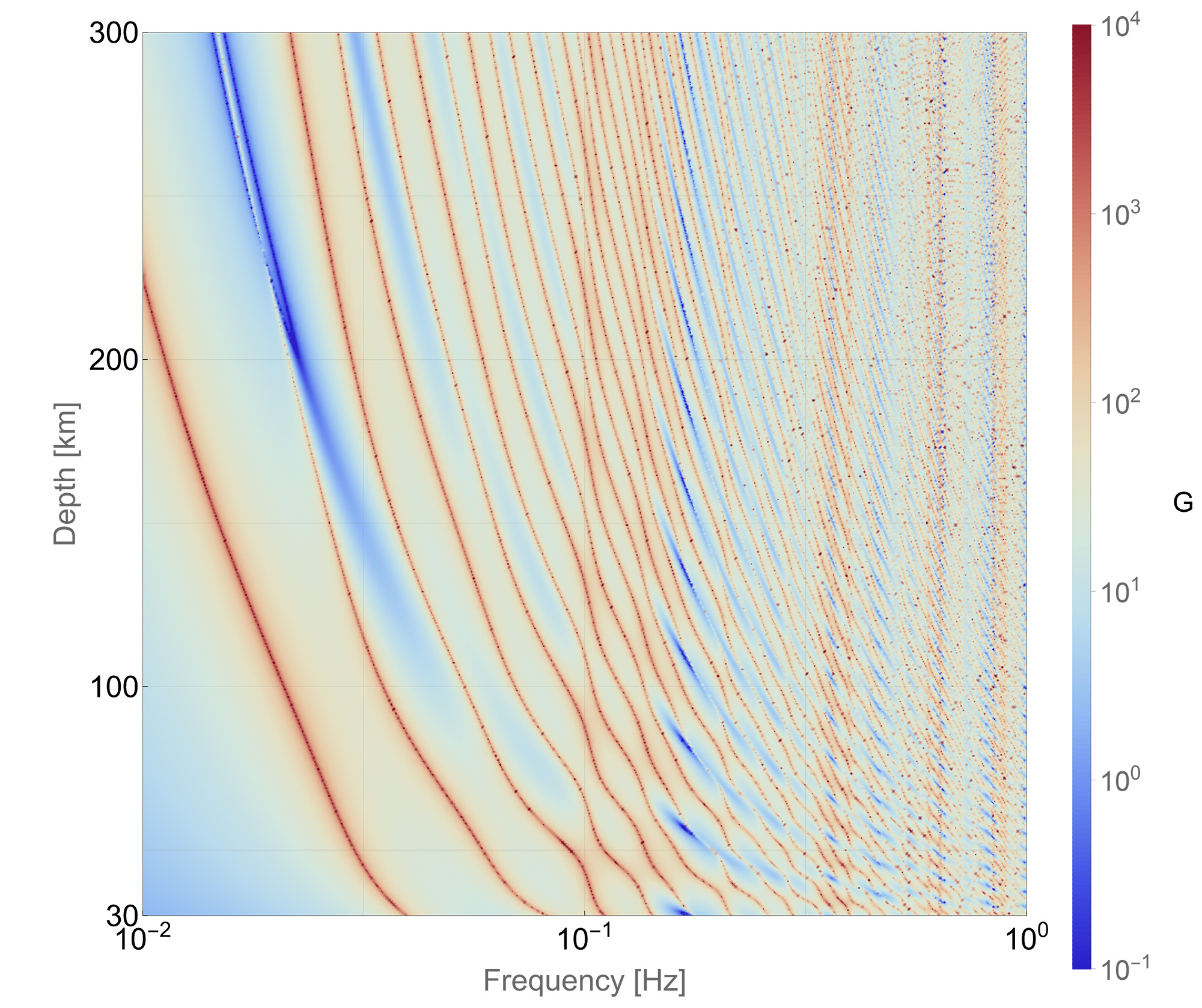}
\caption{Horizontal lunar GW response relative to the homogeneous half-space model with varying depth of an additional layer interface to simulate a strong shear-wave energy return.}
\label{fig:fluidvary}
\end{figure}
The final parametric study assumes an additional layer interface for shear-wave energy return placed at depths from 50\,km to 1000\,km. Technically, we simulate it as a fluid layer, but this is only to simplify the simulation. The resulting response shown in figure \ref{fig:fluidvary} has again the structure with the sharp peaks as in figure \ref{fig:fluidresp}. Also here, the height of the peaks is ultimately limited by the Q-value of the layer materials. No matter how deep the layer, there is a dense series of sharp response peaks in the 0.1\,Hz to 1\,Hz band. The simulation was made with a $Q=2000$ for all layers, but without significant impact on the results compared to an infinitely high Q-value. If the Q-value limitations of the material become important, then they would reduce the response at 1\,Hz more than at 0.1\,Hz, and the effect would depend on how deep the energy-return layer is. 

\section{LGWA sensitivity in the decihertz band}
\label{sec:sensitivity}
\begin{figure}
\centering
    \includegraphics[width=0.99\columnwidth]{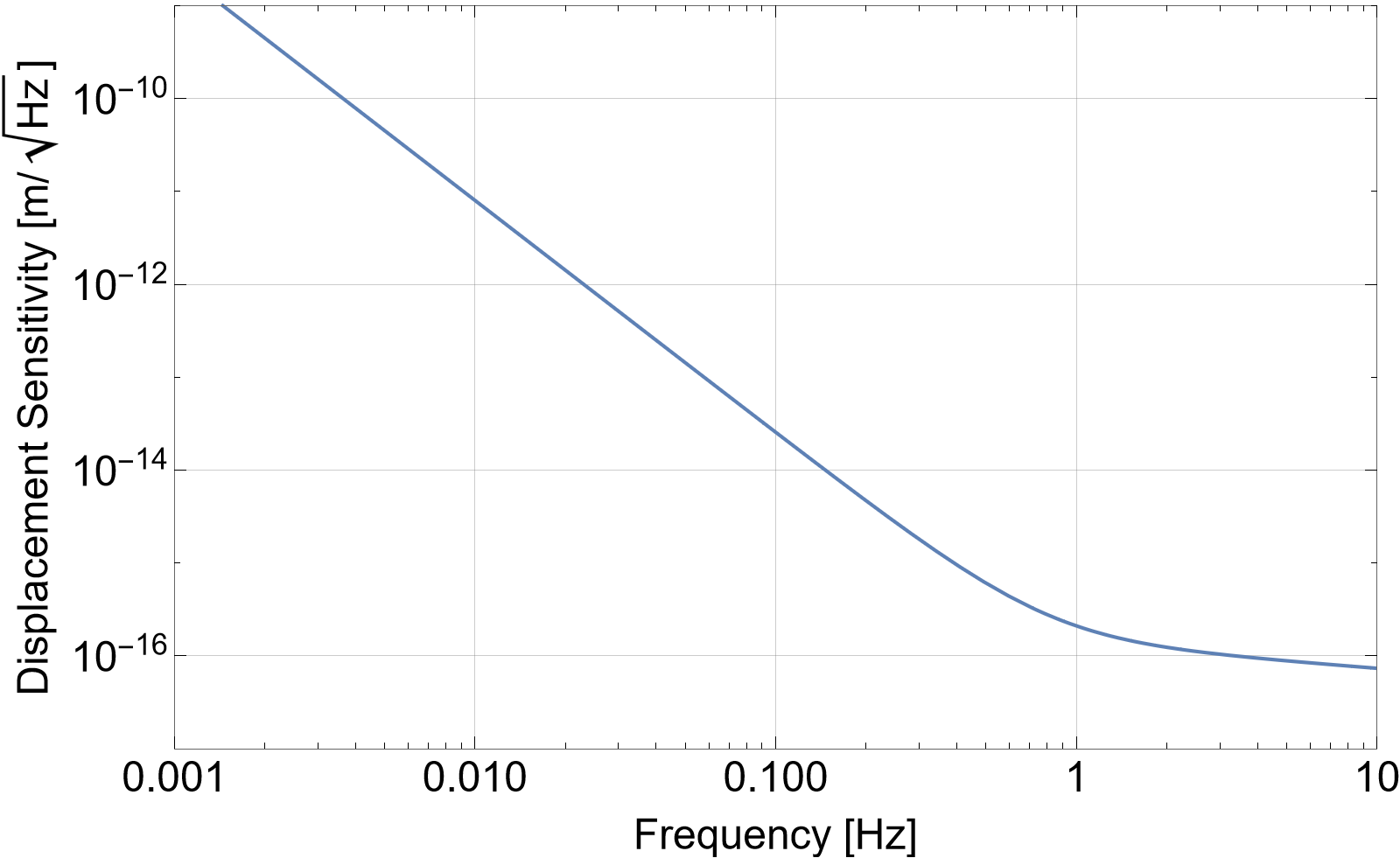}
  \caption{Targeted sensitivity of an LGWA accelerometer to ground displacement.}
\label{fig:sensor}
\end{figure}

We can now use the results of the previous section to calculate the sensitivity of the LGWA. Specifically, we use the results for the GW response $\mathcal{R}(f)$ shown in figure \ref{fig:fluidresp} to convert the displacement sensitivity of an LGWA accelerometer into a GW strain sensitivity. The target displacement sensitivity of the LGWA accelerometers is shown in figure \ref{fig:sensor} as square-root of the power spectral density $S(f)$. 
In this model, the noise below 1\,Hz is dominated by suspension thermal noise of a Watt's linkage assumed here out of silicon while the noise above 1\,Hz is dominated by readout noise of the superconducting coils combined with a superconducting quantum interference device used as low-noise amplifier. The corresponding LGWA sensitivity model as characteristic strain is then calculated as
\begin{equation}
    h_n(f)=\frac{\sqrt{fS(f)}}{2\mathcal R(f)},
\end{equation}
where the factor 2 in the denominator is a standard recipe used for LGWA to account for the fact that the accelerometer array is assumed to consist of 4 stations, and each station measures ground displacement along two orthogonal horizontal directions. This is only an approximate correction and to be accurate, one must consider both GW polarizations and data from all 8 accelerometer channels to calculate the full LGWA sensitivity as for example implemented in GWFish \cite{DuEA2022}.

\begin{figure}
\includegraphics[width=0.99\columnwidth]{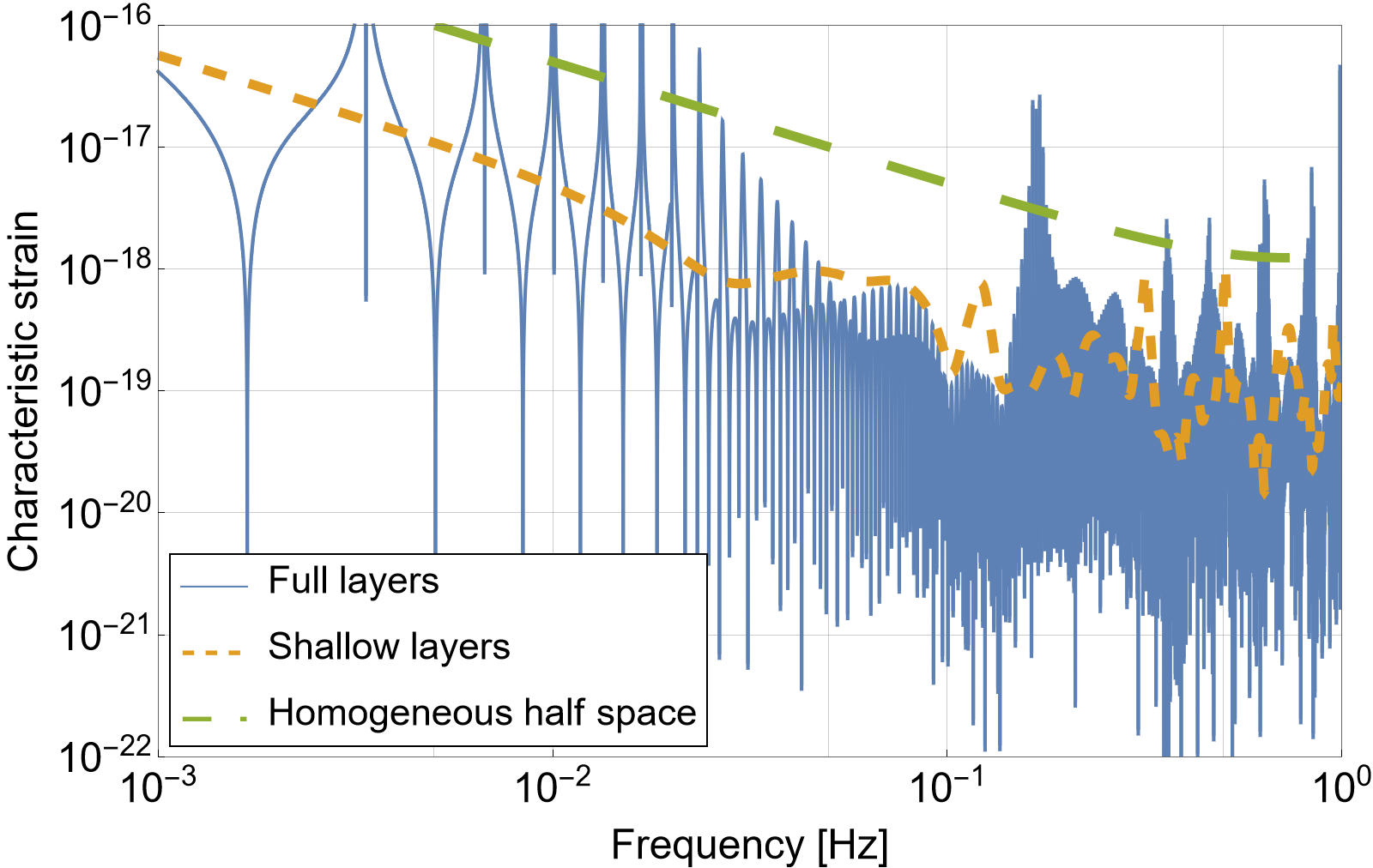}
\includegraphics[width=0.99\columnwidth]{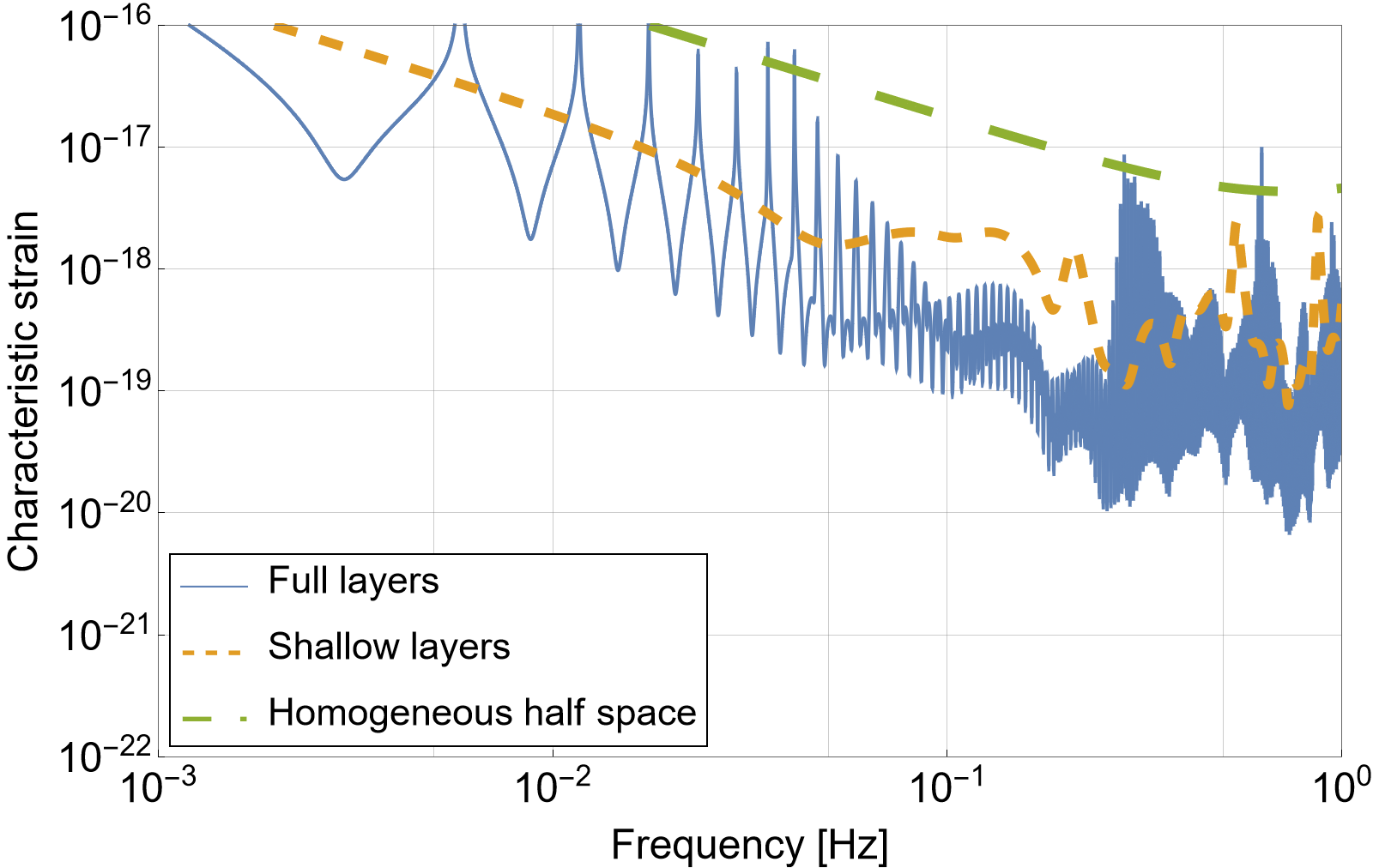}
\caption{GW strain sensitivity using horizontal (upper plot) and vertical (lower plot) accelerometers in the LGWA array. In reality, the difference between the two responses will not be so large, and is a result of how we simulate the energy return in our model, i.e., for these plots through a deep fluid layer.}
\label{fig:sensitivity}
\end{figure}

The results for the GW strain sensitivities are shown in figure \ref{fig:sensitivity} measuring surface displacements along the horizontal (a) and vertical (b) direction. 
The large difference between the two plot comes from our (artificial) choice to simulate the energy return with the help of a fluid layer. As before, this leads to a near-perfect energy storage for shear waves where energy loss is dominated by the damping introduced by the rock itself ($Q=2000$), while the energy loss in the compressional-wave field is dominated by transmission into the fluid layer. 

\section{Discussion}
The reference model presented in section \ref{sec:horizontally} that includes only the shallow geological layers of table \ref{tab:refmodel} does not explain important features of moonquake observations and active studies during the Apollo missions especially concerning the very high Q-values inferred from moonquake coda decay times and envelop fits. A summary of the Q-measurements can be found in table 3 of Garcia et al \cite{GaEA2019}. There are two Q-measurements at frequencies between 0.1\,Hz and 1\,Hz. One of them reports a shear-wave Q-value of 5000 \cite{Dainty1974}, the other of 2500 \cite{Gillet2017}. These Q-values are similar to what was inferred from measurements above 1\,Hz with the lowest reported value of $Q=1600$ for shear waves at 4\,Hz within the first 2\,km of the ground. 

It is important to interpret the Q-values. A major distinction is between absorption and scattering Q-values. Absorption Q-values quantify an actual energy loss, and they have a value of a few 1000 in the upper layers of the Moon \cite{Gillet2017}. Instead, the scattering Q-value is a measure of diffusivity characterizing the wave propagation, and this value is typically low, e.g., less than 100 in the megaregolith down to about 80\,km and then steeply increasing to a few 1000 below 100\,km depth \cite{Gillet2017}. Diffusion does not necessarily mean loss of energy, but rather loss of certain type of information. In fact, based on recent studies of seismic wave propagation on the Moon \cite{menina2023scatt}, simulations suggest that dense fracture networks might be the main reason for the long moonquake coda times. According to these simulations, the layered structure is not essential to explain the long observation times (coda decay times) of moonquakes. Also taking into account that GWs produce seismic waves coherently over the entire Moon, the scattering phenomena work as lunar GW response amplification because they greatly extend the impulse response of a GW signal. An open question is how accurately these fracture networks could be characterized around the LGWA deployment site to allow for an accurate calibration of LGWA data and possibly even retrieving phase information about the GW signal from the seismic signal, which might get lost in a seismic field subject to diffuse scattering. Hence, the lunar GW-response estimates for horizontal displacement measurements provided in this paper are consistent with known properties of the lunar megaregolith, but it is still necessary to find a way to include the fracture networks in the layered model; at least in such a way that their effect on wave propagation is simulated. Until then, important question marks remain concerning the lunar GW response.

It is also important to note that the lunar GW response in the decihertz band is expected to vary significantly across the lunar surface due to topographic features and geological heterogeneity. This is why eventually it will be necessary to switch to full dynamical simulations of the lunar GW response based on finite-element models. Future proposed missions (e.g., Lunar Geophysical Network \cite{Weber+2020}) or missions already under preparation (Farside Seismic Suite and Chang'e 7 \cite{Panning+2022FSS,Chi2023}), which include seismometers in their payloads, might provide important new insight into the properties of the lunar seismic field and the structure of the megaregolith to refine our response models. Especially the proposed LGWA pathfinder mission Soundcheck with deployment site in one of the Moon's permanently shadowed regions would provide crucial information for LGWA mission planning \cite{harms2022research}. 

\section{Conclusions}
In this paper, we present a new formalism to simulate the lunar GW response in a layered geology. The method is an extension of the Dyson model of the GW response of a homogeneous half space. A coordinate system is used where the GW couples with the elastic medium through gradients of the shear modulus. 

The first important result of our study is that the GW response is strongly amplified in multi-layered geologies compared to the homogeneous half-space model. Amplifications greater than a factor 10 are found with respect to the homogeneous model. We then probe the dependence of the results on certain parameter values of the layered model including layer thickness, material Q-value (absorption), and shear-wave speed. We find that low shear-wave speed in the first layer increases the lunar GW response. This is mostly due to the increased shear-modulus change to the next layer and corresponding enhancement of seismic reflection from the first layer interface and corresponding resonant enhancement of seismic waves within the first layer.

The most important change to the results comes when introducing features that return shear-wave energy more effectively. We use a fluid layer as an (artificial) means to create strong reflection of shear waves. No matter how deep this layer is located, it strongly enhances the GW response by orders of magnitude introducing strongly peaked resonances. At lower frequencies, a liquid core or even the free surface at the other side of the Moon can create strong back reflections. 

An important mechanism that influences energy transport in the upper layers of the Moon (the lunar megaregolith) is seismic scattering. It requires a separate study to develop a method to simulate the effect of fracture networks on the lunar GW response. Simulations showed that fracture networks can store seismic energy for long times and release it gradually to produce long moonquake coda characterized by Q-values of a few 1000. How such high-Q phenomena, even if through scattering, influence the lunar GW response will be subject to future studies.

\section{Acknowledgments}
The authors thank Sabrina Menina for helpful discussions about seismic-wave propagation on the Moon.

\bibliography{references}

\end{document}